\documentclass{article}
\usepackage{kotex}
\usepackage{xcolor}
\usepackage{subcaption}
\usepackage{graphicx}
\usepackage{spconf,amsmath,graphicx}
\usepackage{amsfonts}
\usepackage{verbatim}
\usepackage{multirow}
\usepackage[ruled]{algorithm2e}
\usepackage{url}

\title{ATTENTION BASED ON-DEVICE STREAMING SPEECH RECOGNITION \\ WITH LARGE SPEECH CORPUS}
\name{Kwangyoun Kim*, Kyungmin Lee*, Dhananjaya Gowda, Junmo Park, Sungsoo Kim, Sichen Jin, 
      \thanks{* Equal contribution. The author would like to thank Jiyeon Kim, Mehul Kumar and Abhinav Garg for constructive comment.}}
\secondlinename{Young-Yoon Lee, Jinsu Yeo, Daehyun Kim, Seokyeong Jung, Jungin Lee, Myoungji Han, Chanwoo Kim}
\address{Samsung Electronics Co., Ltd., Korea}
\begin{document}
%
\maketitle
\begin{abstract}
In this paper, we present a new on-device automatic speech recognition (ASR) system based on monotonic chunk-wise attention (MoChA) models trained with large ($> 10K$ hours) corpus.
We attained around 90\% of a word recognition rate for general domain mainly by using joint training of connectionist temporal classifier (CTC) and cross entropy (CE) losses, minimum word error rate (MWER) training, layer-wise pre-training and data augmentation methods.
In addition, we compressed our models by more than 3.4 times smaller using an iterative hyper low-rank approximation (LRA) method while minimizing the degradation in recognition accuracy.
The memory footprint was further reduced with 8-bit quantization to bring down the final model size to lower than 39 MB.
For on-demand adaptation, we fused the MoChA models with statistical n-gram models, and we could achieve a relatively 36\% improvement on average in word error rate (WER) for target domains including the general domain.
\end{abstract}
\section{Introduction}
\label{sec:intro}
Recently, End-to-end (E2E) neural network architectures based on sequence to sequence (seq2seq) learning for automatic speech recognition (ASR) have been gaining lots of attention~\cite{Battenberg2017ExploringNT,Prabhavalkar2017}, mainly because they can learn both the acoustic and the linguistic information, as well as the alignments between them, all simultaneously unlike the conventional ASR systems which were based on the hybrid models of hidden Markov models (HMMs) and deep neural network (DNN) models.
Moreover, the E2E models are more suitable to be compressed since they do not need separate phonetic dictionaries and language models, making them one of the best candidates for on-device ASR systems.

Among the various E2E ASR model architectures such as attention-based encoder-decoder models~\cite{Chan2015ListenAA} and recurrent neural network transducer (RNN-T) based models~\cite{Graves2012SequenceTW,Rao2017ExploringAD}, we chose to use the attention based method since the accuracy of this method has surpassed that of the conventional HMM-DNN based state-of-the-art ASR systems~\cite{Chiu2018StateoftheArtSR}.
Despite their extreme accuracy, attention models which require full alignment between the input and the output sequences are not capable of providing streaming ASR services.
Some researches have been made to address this lack of streaming capabilities of the attention models~\cite{45549,Raffel2017OnlineAL,Chiu2018MonotonicCA}.
In \cite{45549}, an online neural transducer was proposed, which applies the full attention method on chucks of input, and is trained with an additional end-of-chunk symbol.
In \cite{Raffel2017OnlineAL}, a hard monotonic attention based model was proposed for streaming decoding with acceptable accuracy degradation.
Furthermore, in ~\cite{Chiu2018MonotonicCA}, a monotonic chunk-wise attention (MoChA) method was proposed, which showed the promising accuracy by loosening a hard monotonic alignment constraint and using a soft attention over a few speech chunks.

In this paper, we explain how we improved our MoChA based ASR system to become an on-device commercialization ready solution.
First, we trained the MoChA models by using connectionist temporal classification (CTC) and cross-entropy (CE) losses jointly to learn alignment information precisely.
A minimum word error rate (MWER) method, which is a type of sequence-discriminative training, was adopted to optimize the models~\cite{Prabhavalkar2018MinimumWE}.
Also, for better stability and convergence of model training, we applied a layer-wise pre-training mechanism~\cite{zeyer2018:attanalysis}.
Furthermore, in order to compress the models, we present a hyper low-rank matrix approximation (hyper-LRA) method by employing DeepTwist~\cite{Lee2015Deeptwist} with minimum accuracy degradation. 
Another important requirement for the commercializing ASR solutions is to boost the recognition accuracy for user context-specific keywords. 
In order to bias the ASR system during inference time, we fused the MoChA models with statistical n-gram based personalized language models (LMs).

The main contribution of this paper is in successfully building the first ever attention-based streaming ASR system capable of running on devices to the best of our knowledge.
We succeeded not only in training MoChA models with large corpus for Korean and English, but also in satisfying the needs of commercial on-device applications.

The rest of this paper is composed as follows: the speech recognition models based on attention methods are explained in a section 2. A section 3 describes how optimization methods improved recognition accuracy, and explanation for the compression algorithm for MoChA models is followed in a section 4. A section 5 describes the n-gram LM fusion for on-demand adaptation, and then discusses the methods and related experiments results in a section 6 and 7.
%
\section{MODEL ARCHITECTURE}
\label{sec:model}
%
Attention-based encoder-decoder models are composed of an encoder, a decoder, and an attention block between the two~\cite{Bahdanau2015Att}.
The encoder converts an input sequence into a sequence of hidden vector representations referred to as encoder embeddings.
The decoder is a generative model which predicts a sequence of the target labels.
The attention is used to learn the alignment between the two sequences of the encoder embeddings and the target labels.

\subsection{Attention-based speech recognition}
\label{sec:bfa}
The attention-based models can be applied to ASR systems~\cite{Chorowski2015Att, returnn_asr} using the following equations (1)-(4).
%
\begin{equation}
h_t = Encoder(h_{t-1}, x_t)
\end{equation}
where $\textbf{x}=\{x_1,x_2,...,x_T\}$ is the speech feature vector sequence, and $\textbf{h}=\{h_1,h_2,...,h_T\}$ is the sequence of encoder embeddings.
The $Encoder$ can be constructed of bi- or uni-directional long short term memory (LSTM) layers~\cite{Hochreiter:1997:LSM:1246443.1246450}. 
Due to the difference in length of the input and the output sequence, the model is often found to have difficulty in convergence. 
In order to compensate for this, pooling along the time-axis on the output of intermediate layers of the encoder is used, effectively reducing the length of \textbf{h} to $T' < T$.
\begin{equation}
\begin{aligned}
\label{eq:attention}
e_{t,l} &= Energy(h_t, s_l)\\
&= v^T\tanh(W_hh_t + W_ss_{l} + b)\\
a_{t,l} &= Softmax(e_{t,l})
\end{aligned}
\end{equation}
An attention weight $a_{t,l}$, which is often referred as alignment, represents how the encoder embeddings of each frame $h_t$ and the decoder state $s_l$ are correlated~\cite{Bahdanau2015Att}.
We employed an additive attention method to compute the correlations. A softmax function converts the attention energy into the probability distribution which is used as attention weight.
A weighted sum of the encoder embeddings is computed using the attention weights as,
\begin{equation}
\begin{aligned}
c_l &= \sum_{t}^{T'}a_{t,l}h_{t}
\label{eq:context}
\end{aligned}
\end{equation}
where $c_l$ denotes the context vector, and since the encoder embeddings of the entire input frames are used to compute the context, we could name this attention method as full attention.
The $Decoder$, which consists of uni-directional LSTM layers, computes the current decoder state $s_l$ from the previous predicted label $y_{l-1}$ and the previous context $c_{l-1}$.
And the output label $y_l$ is calculated by a $Prediction$ block using the current decoder state, the context and the previous output label. 
\begin{equation}
\begin{aligned}
s_l &= Decoder(s_{l-1}, y_{l-1}, c_{l-1})\\
y_l &= Prediction(s_l, y_{l-1}, c_l)
\end{aligned}
\end{equation}

Typically, the prediction block consists of one or two fully connected layers and a softmax layer to generate a probability score for the target labels. We applied max pooling layer between two fully connected layers.
The probability of predicted output sequence $\textbf{y}=\{y_1,y_2,...,y_L\}$ for given $\textbf{x}$ is calculated as in equation (5).
\begin{equation}
\begin{aligned}
P(\textbf{y}|\textbf{x}) = \prod_l^L{P(y_l|\textbf{x}, y_{1:l-1})}
\end{aligned}
\end{equation}
where $P(y_l|\textbf{x}, y_{1:l-1})$ is the probability of each output label.
%
%
Even though the attention-based models have shown state-of-the-art performance, they are not a suitable choice for the streaming ASR, particularly because they are required to calculate the alignment between the current decoder state and the encoder embeddings of the entire input frames.
\subsection{Monotonic Chunk-wise Attention}
\label{sec:mocha}
A monotonic chunk-wise attention (MoChA) model is introduced to resolve the streaming incapability of the attention-based models under the assumption that the alignment relationship between speech input and output text sequence should be monotonic~\cite{Raffel2017OnlineAL, Chiu2018MonotonicCA}.

MoChA model computes the context by using two kinds of attentions, a hard monotonic attention and a soft chunkwise attention. The hard monotonic attention is computed as,
\begin{equation}
\label{eq:MonoAttention}
\begin{aligned}
e_{t,l}^{monotonic} &= MonotonicEnergy(h_t, s_{l})\\
a_{t,l}^{monotonic} &= \sigma(e_{t,l}^{monotonic})\\
z_{t,l} &\sim Bernoulli(a_{t,l}^{monotonic})\\
&= 
\begin{cases}
    1, & \text{if } a_{t,l}^{monotonic} \geq 0.5\\
    0, & \text{otherwise}
\end{cases}
\end{aligned}
\end{equation}
where $z_{t,l}$ is the hard monotonic attention used to determine whether to attend the encoder embedding $h_t$. 
The $Decoder$ attends at $u^{th}$ encoder embedding to predict next label only if $z_{u,l}=1$.
The equation (6) is computed on $t \geq u_{l-1}$, where $u_{l-1}$ denotes the attended encoder embedding index for previous output label prediction.
The soft chunkwise attention is computed as
\begin{equation}
\label{eq:ChunkAttention}
\begin{aligned}
e_{t,l}^{chunk} &= ChunkEnergy(h_t, s_{l})\\
a_{t,l}^{chunk} &= Softmax(e_{t,l}^{chunk})\\
c_{t,l} &= \sum_{t=u-w+1}^{u}a_{t,l}^{chunk}h_{t}
\end{aligned}
\end{equation}
where $u$ is the attending point chosen from monotonic attention, $w$ is the pre-determined chunk size, and $a_{t,l}^{chunk}$ is the chunkwise soft attention weight and $c_{t,l}^{chunk}$ is the chunkwise context which is used to predict the output label.

We used the modified additive attention for computing the attention energy in order to ensure model stability~\cite{Chiu2018MonotonicCA}.
\begin{equation}
Energy'(h_t, s_{l}) =
g\frac{v}{||v||}\tanh(W_hh_t + W_ss_{l-1} + b) + r\\
\end{equation}
where $g$, $r$ are additional trainable scalars, and others are same as $Energy()$ in equation (\ref{eq:attention}). $MonotonicEnergy()$ and $ChunkEnergy()$ are computed using equation (8) with own trainable variables.  
%
%
%
\begin{figure}[h]
  \centering
  \includegraphics[trim={0 0cm 0 0cm}, clip, width=.8\linewidth]{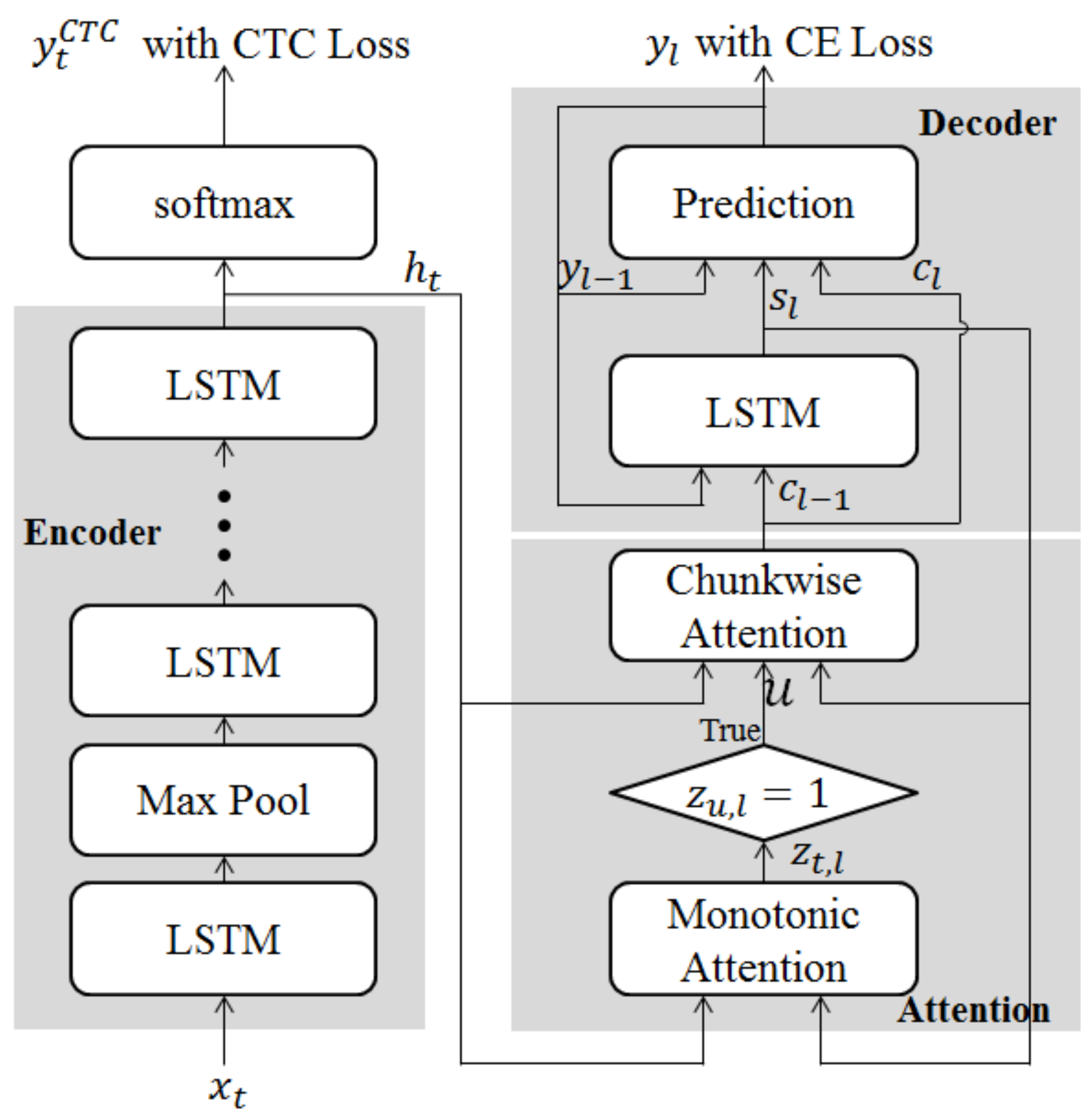}
  \caption{Model architecture}  
\end{figure} 
\section{Training and Optimization}
\label{sec:optimization}
The main objective of the attention-based encoder-decoder model is to find parameters which minimize the the cross entropy (CE) between the predicted sequences and the ground truth sequences.
\begin{equation}
\label{eq:ce}
L_{CE} = -\sum_{l}\log(P(y_l^*|\textbf{x},y_{1:l-1}^*))
\end{equation}
where $y^*$ is the ground truth label sequence. 
We trained MoChA models with CTC loss and CE loss jointly for learning the alignment better and the MWER loss based sequence-discriminative training was employed to further improve the accuracy.
Moreover in order to ensure the stability in training MoChA models, we adopted the pre-training scheme.
\subsection{Joint CTC-CE training}
\label{ssec:ctc-ce}
%
In spite of the different length between the speech feature sequences and the corresponding text sequences, CTC loss induces the model that the total probability of all possible alignment cases between the input and the output sequence is maximized~\cite{Graves2006ConnectionistTC}.
The CTC loss are defined as follows,
\begin{equation}
\label{eq:CTC}
L_{CTC} = -\sum_{\pi\subset\Pi(y^*)}\sum_{t}^{T}\log(P(y_t^\pi|x_t)
\end{equation}
where $\Pi(y^*)$ are all of the possible alignments generated with \{Blank\} symbol and the repetition of outputs units for having same length as input speech frames, and $\pi$ is one of them. $P(y_t^\pi|x_t)$ is the probability about $t$th predicted label is $t$th label in $\pi$ alignment case.

A CTC loss can be readily applicable for training MoChA Model, especially the encoder, because it also leads the alignment between input and output as a monotonic manner. Moreover CTC loss has the advantage of learning alignment in noisy environments and can help to quickly learn the alignment of the attention based model through joint training~\cite{Kim2017JointCB}.

The joint training loss is defined as follows,
\begin{equation}
\label{eq:jointCTCLoss}
L_{Total} = \lambda L_{CE} + (1-\lambda) L_{CTC} \quad\;\lambda \in [0,1]
\end{equation}
where $L_{Total}$ is joint loss of the two losses.
\subsection{MWER training}
\label{ssec:mwer}
%
In this paper, the byte-pair encoding (BPE) based sub-word units were used as the output unit of the decoder ~\cite{DBLP:conf/acl/SennrichHB16a}.
Thus the model is optimized to generate individual BPE output sequences well.  
However, the eventual goal of the speech recognition is to reduce the word-error rate (WER).
Also, since the decoder is used along with a beam search during inference, it is effective to improve the accuracy by directly defining a loss that lowers the exptected WER of candidated beam search results.
The loss of MWER is represented as follows,
\begin{equation}
\label{eq:MWERLoss}
L_{MWER} = \sum_{b \subset B}P(\textbf{y}^b|\textbf{x})(\mathcal{W}^b-\mathcal{\Bar{W}})  
\end{equation}
where $B$ are all the candidates of beams search results, and $\mathcal{W}^b$ indicates the number of word error of each beam result sequence $y^b$. The average word error of all the beam $\mathbb{\Bar{W}}$ helps model converging well by reducing the variance.
\begin{equation}
\label{eq:jointMWERLoss}
L_{Total} = \lambda L_{CE} + (1-\lambda) L_{MWER} \quad\;\lambda \in [0,1]
\end{equation}
The MWER loss, $L_{MWER}$, also can be easily integrated with the CE loss by linearly interpolating as shown in Equation~\ref{eq:jointMWERLoss}.
%
%
\subsection{Layer-wise pre-training}
\label{ssec:pretrain}
%
A layer-wise pre-training of the encoder was proposed in \cite{zeyer2018:attanalysis} to ensure that the model converges well and has better performance. 
The initial encoder consists of 2 LSTM layers with a max-pool layer with a pooling factor of 32 in between. After sub-epochs of training, a new LSTM layer and a max-pool layer are added. 
The total pooling factor 32 is divided into 2 for lower layer and 16 for newly added higher layer. 
This strategy is repeated until the entire model network is completely built with 6 encoder LSTM layers and 5 max-pool layers with 2 pooling factor. 
Finally, the total reducing factor is changed to 8, with only the lower 3 max-pool layers having a pooling factor 2.

During pre-training of our MoChA models, when a new LSTM and a max-pool layer were piled up at each stage, the training and validation errors shot up.
In order to address this, we employed a learning rate warm-up for every new pre-training stage.
%
\subsection{Spec augmentation}
\label{ssec:augmentation}
Because end-to-end ASR model learns from the transcribed corpus, large dataset is one of the most important factor to achieve better accuracy. 
Data augmentation methods have been introduced to generate additional data from the originals, and recently, spec augmentation shows state-of-the-art result on public dataset~\cite{spec2019}. 
spec augmentation masks parts of spectrogram along the time and frequency axis, thus model could learn from masked speech in a lack of information. 
\section{Low-rank matrix approximation}
\label{sec:LRA}
%
We adopted a low-rank matrix approximation (LRA) algorithm based on singular value decomposition (SVD) to compress our MoChA model~\cite{Xue2014SVD}.
Given a weight matrix $W \in \mathbb{R}^{m \times n} $, SVD is $U\Sigma V^T $, where $\Sigma\in \mathbb{R}^{m \times n}$ is a diagonal matrix with singular values, and $U\in \mathbb{R}^{m\times m}$ and $V\in \mathbb{R}^{n \times n}$ are unitary matrices.
If we specify the rank $r< \frac{mn}{m+n}$, the truncated SVD is given by  $\tilde{W}=\tilde{U}\tilde{\Sigma}\tilde{V}^T \in \mathbb{R}^{m\times n}$, where $\tilde{U}\in \mathbb{R}^{m\times r},  \tilde{\Sigma}\in \mathbb{R}^{r\times r}$, and $\tilde{V}\in \mathbb{R}^{n\times r}$ are the top-left submatrices of $U,\Sigma$, and $V$, respectively.
For an LRA, $W$ is replaced by the multiplication of $\tilde{U}'=\tilde{U}\tilde{\Sigma}$ and $\tilde{V}^T$, the number of weight parameters is reduced as $r(m+n) < mn$.
Hence we obtain the compression ratio $\rho=\frac{mn}{r(m+n)}$ 
in memory footprints and computation complexity for matrix-vector multiplication.
From LRA, we have an LRA distortion as
\begin{align}
\label{eq:lra_dist}
    \Delta W = \tilde{U}' \tilde{V}^T - W.
\end{align}
For each layer, given an input $x \in \mathbb{R}^m$, the output error is given by
\begin{equation}
\label{eq:lra}
e = \sigma((W+\Delta W) x +b)-\sigma(W x +b), 
\end{equation}
where $b\in \mathbb{R}^n$ and $\sigma(\cdot)$ represent a bias vector and a non-linear activation function, respectively.
Then it propagates through the layers and increases the training loss.
In the LRA, $\tilde{U}'$ and $\tilde{V}^T$ are updated in backward pass by constraining the weight space of $r(m+n)$ dimensions.
However, for large $\rho$, it is difficult to find the optimal weights due to the reduced dimension of the weight space.
Instead, we find an optimal LRA by employing DeepTwist method~\cite{Lee2015Deeptwist}, called a hyper-LRA.
\begin{algorithm}[htb]
\SetAlgoLined
\SetKwFunction{PTrain}{TrainingModelWeights}
\SetKwFunction{PInfer}{InferenceModelWeights}
\SetKwProg{Fn}{Procedure}{}{}
\Fn{\PTrain{}}{
 \KwResult{Weight matrices $\{ W_i \}$ }
 \For{each iteration $N$}{
  \For{each layer $i$}{
  $x_{i+1} \gets \sigma(W x_i +b)$
  \tcp*{$x_i$:\footnotesize\texttt{input}}
  \If{$N \bmod D = 0 $ 
  }{  
   $x_{i+1} \gets x_{i+1} + e_{i}$
   \tcp*{$e_{i}$ \footnotesize\texttt{in (\ref{eq:lra})}} 
   $W_i \gets W_i + \Delta W_i$
   \tcp*{$\Delta W_{i}$ \footnotesize\texttt{in (\ref{eq:lra_dist})}} 
   }{
  }
  } 
  compute the loss $L$\;
 \For{each layer $i$}{
   $W_i \gets W_i - \eta \frac{\partial L}{\partial W_i}$
   \tcp*{$\eta$:\footnotesize\texttt{learning rate}}
 }   
 }
 \For{each layer $i$}{
   $W_i \gets W_i + \Delta W_i$\;
 } 
}
\Fn{\PInfer{}}{
\KwResult{Weight matrices $\{ \tilde{U_i}' \}, \{ \tilde{V_i}^T\}$}
 \For{each layer $i$}{
$\tilde{U_i}',\tilde{V_i}^T \gets truncatedSVD(W_i) $ 
}
}
 \caption{The hyper-LRA algorithms}
\end{algorithm}
The hyper-LRA algorithm modifies retraining process by adding the LRA distortion to weights and the corresponding errors to the outputs of layers every $D$ iterations, where $D$ is a distortion period. After retraining, instead of $W$, the multiplication of $\tilde{U}'$ and $\tilde{V}^T$ is used for the inference model.

Note that the hyper-LRA algorithm optimizes $W$ rather than $\tilde{U}'$ and $\tilde{V}^T$. 
In other words, a hyper-LRA method performs weight optimization in the hyperspace of the truncated weight space, which has the same dimension with the original weight space.
Therefore the hyper-LRA approach can provide much higher compression ratio than the conventional LRA whereas it requires more computational complexity for retraining process.

\section{On-Demand Adaptation}
\label{sec:adaptation}

The on-demand adaptation is an important requirement not only for personal devices such as mobiles but also for home appliances such as televisions.
We adopted a shallow fusion~\cite{kannan_wu_nguyen_sainath_chen_prabhavalkar_2018} method incorporated with n-gram LMs at inference time. 
By interpolating both general LM and domain-specific LMs, we were able to boost the accuracy of the target domains while minimizing degradation in that of the general domain.
The probabilities computed from the LMs and the E2E models are interpolated at each beam step as follows,
\begin{equation}
\begin{aligned}
\label{eq:sf}
P'(y_l|\textbf{x},y_{1:l-1}) &= \log P(y_l|\textbf{x},y_{1:l-1}) \\ &+ \sum_{n=1}^N \alpha_n\log P_{LM_n}(y_l|y_{1:l-1})
\end{aligned}
\end{equation}
where $N$ is the number of n-gram LMs, $P_{LM}$ is a posterior distribution computed by the n-gram LMs. The LM distribution was calculated by looking up a probability per each BPE for the given context.



\section{Experiment}
\label{sec:evlaution}

\subsection{Experimental setup}
\label{ssec:setup}

We evaluated with Librispeech corpus which consists of 960 hours of data first and Internal usage data as well.
The usage corpus consists of around 10K hours of transcribed speech for Korean and English each and was recorded in mobiles and televisions.
We used randomly sampled one hour of usage data as our validation sets for each language.
We doubled speech corpus by adding the random noise both for training and for validating.
The decoding speed evaluated on Samsung Galaxy S10+ equipped with Samsung Exynos 9820 chipsets, a Mali-G76 MP12 GPU and 12GB of DRAM memory.

A sample rate of speech data was 16kHz and the bits per sample were 16.
The speech data were coded with 40-dimensional power mel-frequency filterbank features which are computed by power operation of mel-spectrogram with $\frac{1}{15}$ as the power function coefficient~\cite{vtlp2019}. The frames were computed every 10ms and were windowed by 25ms Hanning window. 
We split words into 10K word pieces through byte pair encoding (BPE) method for both Korean and English normalized text corpus. Especially for Korean, we reduced the total number of units for Korean characters from 11,172 to 68 by decomposing each Korean character into two or three graphemes depending on the presence of final consonants.

We constructed our ASR system based on ReturNN~\cite{returnn}.
In order to speed up the training, we used a multiple GPU training based on the Horovod ~\cite{sergeev2018horovod,Infra2019} all reduce method.
And for better convergence of the model, a ramp-up strategy for both the learning rate and the number of workers was used~\cite{Chiu2018StateoftheArtSR}.
We used a uniform label smoothing method on the output label distribution of the decoder for regularization, and scheduled sampling was applied at a later stage of training to reduce the mismatch between training and decoding. The initial total pooling factor in Encoder is 32 for Librispeech, but 16 is used for internal usage data due to the training sensitivity, and they reduced into 8 after the pre-training stage.
The n-gram LMs were stored in a const arpa structure~\cite{Povey_ASRU2011}.

\subsection{Performance}
\label{ssec:performance}

\begin{table*}[th]
    \caption{The performance of Attention based model depending on the number of direction and the cell size of LSTM layers in encoder. Joint CTC and Label smoothing are applied for all the results, and Data augmentation is only used on Usage data. The beam size of beam search based decoding is 12.}
    \label{tab:mocha_vs_bfa}
    \centering
    \begin{tabular}{cccccccccc}
    \hline
    \multirow{2}{*}{Encoder} & \multirow{2}{*}{Attention} & \multirow{2}{*}{Cell size} 
    &  \multicolumn{2}{c}{Librispeech}
    &  \multicolumn{1}{c}{Usage KOR}
    &  \multicolumn{1}{c}{Usage ENG} \\
    & & &WER(Test-clean) &Test-other & WER &  WER  \\ \hline \hline
    Bi-LSTM  & Full  &  1024    & 4.38\% & 14.34\% & 8.58\% & 8.25\% \\ \hline
    \multirow { 3}{*}{Uni-LSTM} & Full  &  1536    & 6.27\% & 18.42\% & - & - \\
     & \multirow {2}{*}{MoChA} &  1024    & 6.88\% & 19.11\% & 11.34\% & 10.77\% \\
     & &  1536    & 6.30\%  & 18.41\% & 9.33\% & 8.82\% \\ \hline
    \end{tabular}
\end{table*}

\begin{table}[th]
\vspace{-0.2cm}
    \caption{Accuracy improvement from the optimizations}
    \label{tab:trainig_opt}
    \centering
    \begin{tabular}{ccc}
    \hline
     & \multicolumn{2}{c}{Librispeech}\\
     & Test-clean & Test-other \\ \hline \hline
    MoChA (baseline) & 6.70\% & 18.86\%  \\ \hline
     + Joint CTC & \multirow{2}{*}{6.30\%} & \multirow{2}{*}{18.41\%} \\ 
     \& Label smoothing & & \\
     + Spec augmentation & 5.93\% & 15.98\% \\
     + Joint MWER & 5.60\% & 15.52\% \\ \hline
    \end{tabular}
\vspace{-0.2cm}
\end{table}

We performed several experiments to build the baseline model on each dataset, and evaluated accuracies are shown in Table~\ref{tab:mocha_vs_bfa}. In the table, $Bi-$ and $Uni-$ mean bi-directional and uni-directional respectively, and $Cell size$ denotes the size of the encoder LSTM cells. The size of attention dimension is same as the encoder cell size, and 1000 was used the size of the decoder.
The chunk size of MoChA is 2 for all the experiments since we could not see any significant improvement in accuracy by extending the size more than two.
\begin{figure}[h]
  \centering
  \begin{subfigure}{\linewidth}
    \centering
    \includegraphics[trim={0 4.2cm 0 4.2cm}, clip, height=40pt, width=0.8\linewidth]{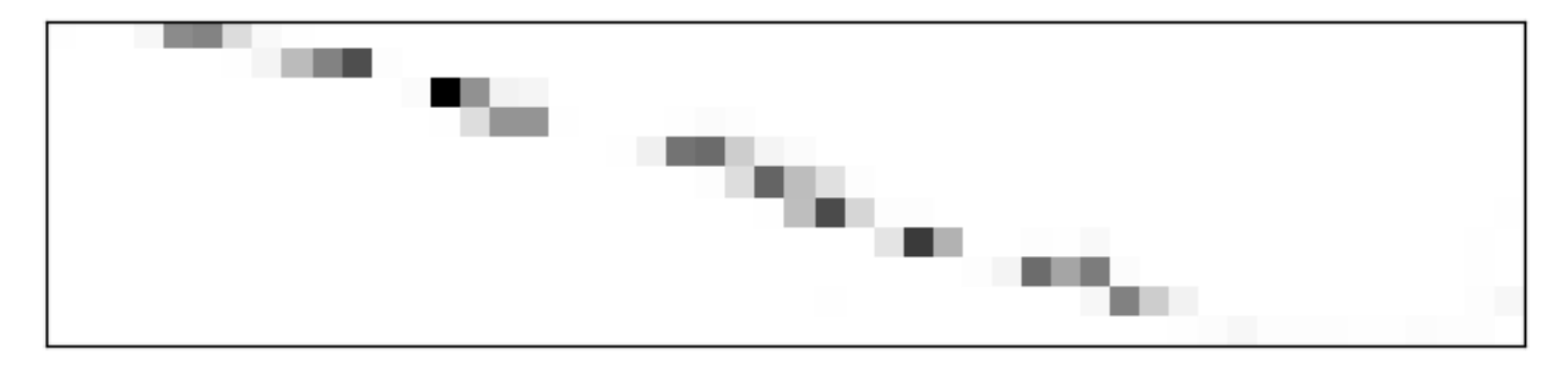}
    \caption{Bi-LSTM Full Attention}
  \end{subfigure}

  \begin{subfigure}{\linewidth}
    \centering
    \includegraphics[trim={0 4.2cm 0 4.2cm}, clip, height=40pt, width=.8\linewidth]{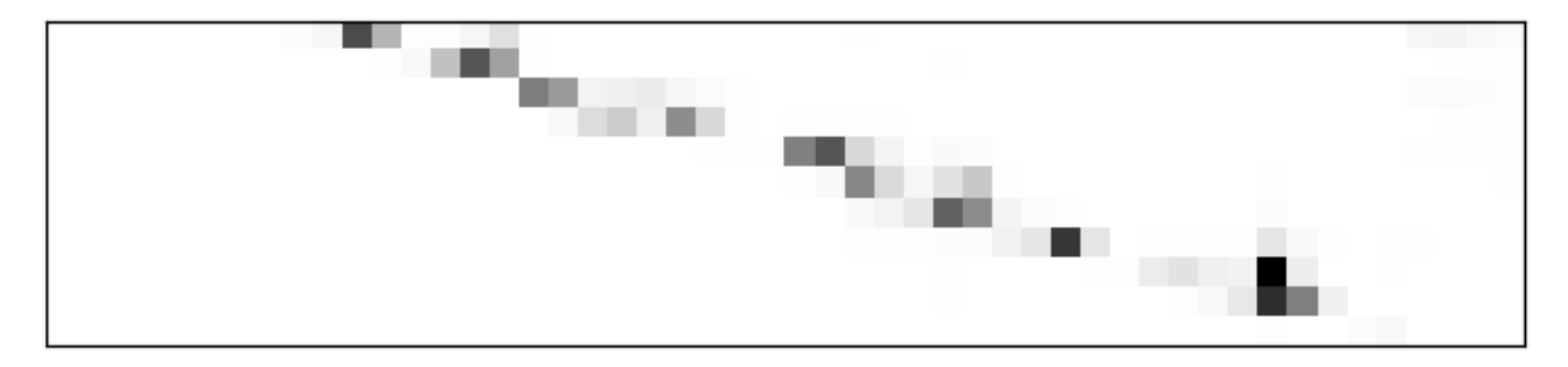}
    \caption{Uni-LSTM Full Attention}
  \end{subfigure}  
  
  \begin{subfigure}{\linewidth}
    \centering
    \includegraphics[trim={0 4.2cm 0 4.2cm}, clip, height=40pt, width=.8\linewidth]{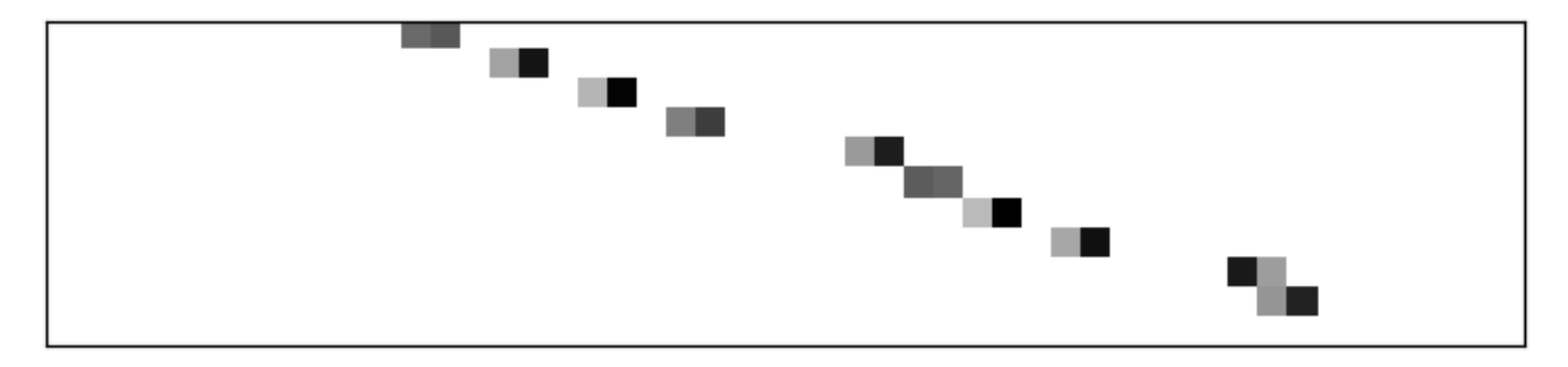}
    \caption{Uni-LSTM MoChA}
  \end{subfigure}  
  \caption{Comparison of alignment by each attention method}  
\end{figure} 

As shown in Fig. 2, compared with bi-directional LSTM case, it seems that uni-directional model's alignment has some time delay because uni-directional LSTM cannot use backward information from input sequence. \cite{Sak2015FastAA}
Alignment calculation with soft full attention may have the advantage of seeing more information and utilizing better context. But for speech recognition, since the alignment of speech utterances is monotonic, that advantage could be not so great.

The accuracy of each trained model with various optimization method are shown in Table~\ref{tab:trainig_opt}. 
The joint weight $\lambda$ for joint CTC training was 0.8, and it was gradually increased during the training.
We used 13 and 50 for the max size of the frequency-axis mask and the time(frame)-axis mask, respectively, and masks were applied one for each. For joint MWER training, we used 0.6 as $\lambda$ and beam size is 4.
Spec augmentation made large improvement, especially on test-other, and after Joint MWER training, finally the accuracies in WERs on test-clean and test-other were improved relatively 16.41\% and 17.71\% respectively compared to that of the baseline.

\subsection{Compression}
\label{ssec:com_result} 
We respectively applied hyper-LRA to weight matrix on each layer,
the rank $r$ of which was chosen empirically. 
For encoder LSTM layers, the ranks of the first and the last layers are set larger than those of the internal layers due to the severity of accuracy degradation. 
The distortion period $D$ was set as the total iterations in one sub-epoch divided by 16.
The compressed model was retrained with a whole training data.
In addition, we adopted 8-bit quantization both to compress and to speed up by using Tensorflow-lite. ~\cite{tensorflow2015-whitepaper, tflite}.
As shown in Table~\ref{tab:compress}, the sizes of the models were reduced at least 3.4 times by applying hyper-LRA, and totally more than 13.68 times reduced after 8-bit quantization with minimum degradation of the performance.
Furthermore, we were able to compensate the performance by using MWER joint training.
At the same time, the decoding speed of Korean and English models got 13.97 and 9.81 times faster than that of baseline models, respectively. The average latency of final models were 140ms and 156ms, and the memory usage during decoding (CODE + DATA) were 230MB and 235MB for Korean and English, respectively.

\begin{table}[th]
\vspace{-0.2cm}
    \caption{Performance for hyper-LRA. 
    The size of models were evaluated in megabytes (MB), and the beam size was 4. xRT denotes real-time factor for decoding speed. }
    \label{tab:compress}
    \centering
    \renewcommand{\tabcolsep}{1.4mm}
    \begin{tabular}{cc|ccc|ccc}
    \hline
    \multirow{2}{*}{Bits} & Hyper &
    \multicolumn{3}{c|}{Korean} & \multicolumn{3}{c}{English} \\
    & LRA & WER & xRT & Size & WER & xRT & Size \\ \hline \hline
    32   &  no                  & 9.37  & 4.89 & 530.56 & 9.03 & 4.32 & 530.50 \\
    32   &  yes                 & 9.85  & 0.99 & 140.18 & 8.91 & 1.15 & 153.98 \\
    32   & +MWER                & 9.60  & 1.26 & 140.18 & 8.64 & 1.48 & 153.98 \\
    8    &  no                  & 9.64  & 1.18 & 132.88 & 9.07 & 0.94 & 132.87 \\
    8    &  yes                 & 10.21 & 0.33 & 35.34  & 9.24 & 0.38 & 38.77 \\
    8    & +MWER                & 9.80  & 0.35 & 35.34  & 8.88 & 0.44 & 38.77 \\ \hline
    \end{tabular}
\vspace{-0.2cm}
\end{table}

\subsection{Personalization}
\label{ssec:pdss}
We evaluated our on-demand adaptation method for the three domains in Korean.
Names of contacts, IoT devices and applications were used to manipulate utterances with pattern sentences where the names were replaced with a class name like "call @name".
Individual n-gram LMs were built for each domain using the synthesized corpus.
LMs for the specific domains were built within 5 seconds as in Table.~\ref{tab:lmbuild}.

\begin{table}[th]
\vspace{-0.2cm}
    \caption{Building times for n-gram LMs (in seconds).}
    \label{tab:lmbuild}
    \centering
    \begin{tabular}{ccccc}
    \hline
    Domain & entities & patterns & utterances & time\\ \hline \hline
    Contact  & 2307 & 23 & 53061 & 4.37 \\ 
    App  & 699 & 25 & 17475 & 1.78 \\ 
    IoT  & 441 & 4 & 1764 & 0.74 \\ \hline
    \end{tabular}
\vspace{-0.2cm}
\end{table}

As in Table~\ref{tab:pdss}, the WER for an App domain was dramatically dropped from 12.76\% to 6.78\% without any accuracy degradation in a general domain.
The additional xRT for the LM fusion was less than 0.15xRT on average even though the number of LM look-up reached millions. The LM sizes for general and for all the three domains were around 43MB and 2MB respectively.
All test sets were recorded in mobiles.

\begin{table}[th]
\vspace{-0.2cm}
    \caption{Performance improvement of on-demand adaptation. *xRTs were evaluated on-devices, but WERs were evaluated on servers with the uncompressed MoChA model in Table~\ref{tab:mocha_vs_bfa}.}
    \label{tab:pdss}
    \centering
    \begin{tabular}{cccccc}
    \hline
    \multirow{2}{*}{Domain} &Length & \multicolumn{2}{c}{MoChA}  &  \multicolumn{2}{c}{Adapted} \\
     &(in hours)&  WER & xRT & WER & xRT \\ \hline \hline
    General &1.0& 9.33  & 0.35 & 9.30 & 0.61 \\
    Contact &3.1& 15.59 & 0.34 & 11.08 & 0.42 \\ 
    App     &1.2& 12.76 & 0.34 & 6.78 & 0.48 \\ 
    IoT     &1.5& 38.83 & 0.43 & 21.92 & 0.52 \\ \hline
    \end{tabular}
\vspace{-0.2cm}
\end{table}

\section{Discussion}
\label{sec:foot}
We accomplished to construct the first on-device streaming ASR system based on MoChA models trained with large corpus.
In spite of the difficulties in training the MoChA models, we adopted various training strategies such as joint loss training with CTC and MWER, layer-wise pre-training and data augmentation.
Moreover, by introducing hyper-LRA, we could reduce the size of our MoChA models to be fit on devices without sacrificing the recognition accuracies.
For personalization, we used shallow fusion method with n-gram LMs, it showed improved results on target domains without sacrificing accuracy for a general domain.
\bibliographystyle{IEEEbib}
\bibliography{strings,refs}

\begin{thebibliography}{10}

\bibitem{Battenberg2017ExploringNT}
Eric Battenberg, Jitong Chen, Rewon Child, Adam Coates, Yashesh Gaur, Yi~Li,
  Hairong Liu, Sanjeev Satheesh, David Seetapun, Anuroop Sriram, and Zhenyao
  Zhu,
\newblock ``Exploring neural transducers for end-to-end speech recognition,''
\newblock {\em 2017 IEEE Automatic Speech Recognition and Understanding
  Workshop (ASRU)}, pp. 206--213, 2017.

\bibitem{Prabhavalkar2017}
Rohit Prabhavalkar, Kanishka Rao, Tara~N. Sainath, Bo~Li, Leif Johnson, and
  Navdeep Jaitly,
\newblock ``A comparison of sequence-to-sequence models for speech
  recognition,''
\newblock in {\em Proc. Interspeech 2017}, 2017, pp. 939--943.

\bibitem{Chan2015ListenAA}
William Chan, Navdeep Jaitly, Quoc~V. Le, and Oriol Vinyals,
\newblock ``Listen, attend and spell,''
\newblock {\em ArXiv}, vol. abs/1508.01211, 2015.

\bibitem{Graves2012SequenceTW}
Alex Graves,
\newblock ``Sequence transduction with recurrent neural networks,''
\newblock {\em ArXiv}, vol. abs/1211.3711, 2012.

\bibitem{Rao2017ExploringAD}
Kanishka Rao, Hasim Sak, and Rohit Prabhavalkar,
\newblock ``Exploring architectures, data and units for streaming end-to-end
  speech recognition with rnn-transducer,''
\newblock {\em 2017 IEEE Automatic Speech Recognition and Understanding
  Workshop (ASRU)}, pp. 193--199, 2017.

\bibitem{Chiu2018StateoftheArtSR}
Chung-Cheng Chiu, Tara~N. Sainath, Yonghui Wu, Rohit Prabhavalkar, Patrick
  Nguyen, Zhifeng Chen, Anjuli Kannan, Ron~J. Weiss, Kanishka Rao, Ekaterina
  Gonina, Navdeep Jaitly, Bo~Li, Jan Chorowski, and Michiel Bacchiani,
\newblock ``State-of-the-art speech recognition with sequence-to-sequence
  models,''
\newblock {\em 2018 IEEE International Conference on Acoustics, Speech and
  Signal Processing (ICASSP)}, pp. 4774--4778, 2018.

\bibitem{45549}
Navdeep Jaitly, David Sussillo, Quoc~V. Le, Oriol Vinyals, Ilya Sutskever, and
  Samy Bengio,
\newblock ``A neural transducer,''
\newblock {\em ArXiv}, vol. abs/1511.04868, 2016.

\bibitem{Raffel2017OnlineAL}
Colin Raffel, Thang Luong, Peter~J. Liu, Ron~J. Weiss, and Douglas Eck,
\newblock ``Online and linear-time attention by enforcing monotonic
  alignments,''
\newblock in {\em ICML}, 2017.

\bibitem{Chiu2018MonotonicCA}
Chung-Cheng Chiu and Colin Raffel,
\newblock ``Monotonic chunkwise attention,''
\newblock in {\em International Conference on Learning Representations}, 2018.

\bibitem{Prabhavalkar2018MinimumWE}
Rohit Prabhavalkar, Tara~N. Sainath, Yonghui Wu, Patrick Nguyen, Zhifeng Chen,
  Chung-Cheng Chiu, and Anjuli Kannan,
\newblock ``Minimum word error rate training for attention-based
  sequence-to-sequence models,''
\newblock {\em 2018 IEEE International Conference on Acoustics, Speech and
  Signal Processing (ICASSP)}, pp. 4839--4843, 2018.

\bibitem{zeyer2018:attanalysis}
Albert Zeyer, André Merboldt, Ralf Schlüter, and Hermann Ney,
\newblock ``A comprehensive analysis on attention models,''
\newblock in {\em Interpretability and Robustness in Audio, Speech, and
  Language (IRASL) Workshop, Conference on Neural Information Processing
  Systems (NeurIPS)}, Montreal, Canada, Dec. 2018.

\bibitem{Lee2015Deeptwist}
D.~Lee, P.~Kapoor, and B.~Kim,
\newblock ``Deeptwist: Learning model compression via occasional weight
  distortion,''
\newblock {\em ArXiv}, vol. abs/1810.12823, 2018.

\bibitem{Bahdanau2015Att}
Dzmitry Bahdanau, Kyunghyun Cho, and Yoshua Bengio,
\newblock ``Neural machine translation by jointly learning to align and
  translate,''
\newblock in {\em 3rd International Conference on Learning Representations,
  {ICLR} 2015, San Diego, CA, USA, May 7-9, 2015, Conference Track
  Proceedings}, 2015.

\bibitem{Chorowski2015Att}
Jan Chorowski, Dzmitry Bahdanau, Dmitriy Serdyuk, Kyunghyun Cho, and Yoshua
  Bengio,
\newblock ``Attention-based models for speech recognition,''
\newblock in {\em Advances in Neural Information Processing Systems 28: Annual
  Conference on Neural Information Processing Systems 2015, December 7-12,
  2015, Montreal, Quebec, Canada}, 2015, pp. 577--585.

\bibitem{returnn_asr}
Albert Zeyer, Kazuki Irie, Ralf Schlüter, and Hermann Ney,
\newblock ``Improved training of end-to-end attention models for speech
  recognition,''
\newblock in {\em Proc. Interspeech 2018}, 2018, pp. 7--11.

\bibitem{Hochreiter:1997:LSM:1246443.1246450}
Sepp Hochreiter and J\"{u}rgen Schmidhuber,
\newblock ``Long short-term memory,''
\newblock {\em Neural Computation}, vol. 9, no. 8, pp. 1735--1780, Nov. 1997.

\bibitem{Graves2006ConnectionistTC}
Alex Graves, Santiago Fern{\'a}ndez, Faustino~J. Gomez, and J{\"u}rgen
  Schmidhuber,
\newblock ``Connectionist temporal classification: labelling unsegmented
  sequence data with recurrent neural networks,''
\newblock in {\em ICML}, 2006.

\bibitem{Kim2017JointCB}
Suyoun Kim, Takaaki Hori, and Shinji Watanabe,
\newblock ``Joint ctc-attention based end-to-end speech recognition using
  multi-task learning,''
\newblock {\em 2017 IEEE International Conference on Acoustics, Speech and
  Signal Processing (ICASSP)}, pp. 4835--4839, 2017.

\bibitem{DBLP:conf/acl/SennrichHB16a}
Rico Sennrich, Barry Haddow, and Alexandra Birch,
\newblock ``Neural machine translation of rare words with subword units,''
\newblock in {\em Proceedings of the 54th Annual Meeting of the Association for
  Computational Linguistics, {ACL} 2016, August 7-12, 2016, Berlin, Germany,
  Volume 1: Long Papers}, 2016.

\bibitem{spec2019}
Daniel~S. Park, William Chan, Yu~Zhang, Chung-Cheng Chiu, Barret Zoph, Ekin~D.
  Cubuk, and Quoc~V. Le,
\newblock ``{SpecAugment: A Simple Data Augmentation Method for Automatic
  Speech Recognition},''
\newblock in {\em Proc. Interspeech 2019}, 2019, pp. 2613--2617.

\bibitem{Xue2014SVD}
J.~Xue, J.~Li, D.~Yu, M.~Seltzer, and Y.~Gong,
\newblock ``Singular value decomposition based low-footprint speaker adaptation
  and personalization for deep neural network,''
\newblock in {\em ICASSP}, 2014.

\bibitem{kannan_wu_nguyen_sainath_chen_prabhavalkar_2018}
Anjuli Kannan, Yonghui Wu, Patrick Nguyen, Tara~N. Sainath, Zhijeng Chen, and
  Rohit Prabhavalkar,
\newblock ``An analysis of incorporating an external language model into a
  sequence-to-sequence model,''
\newblock {\em 2018 IEEE International Conference on Acoustics, Speech and
  Signal Processing (ICASSP)}, 2018.

\bibitem{vtlp2019}
Chanwoo Kim, Minkyu Shin, Abhinav Garg, and Dhananjaya Gowda,
\newblock ``{Improved Vocal Tract Length Perturbation for a State-of-the-Art
  End-to-End Speech Recognition System},''
\newblock in {\em Proc. Interspeech 2019}, 2019, pp. 739--743.

\bibitem{returnn}
Patrick Doetsch, Albert Zeyer, Paul Voigtlaender, Ilya Kulikov, Ralf Schlüter,
  and Hermann Ney,
\newblock ``Returnn: The rwth extensible training framework for universal
  recurrent neural networks,''
\newblock in {\em 2017 IEEE International Conference on Acoustics, Speech and
  Signal Processing (ICASSP)}, 2017.

\bibitem{sergeev2018horovod}
Alexander Sergeev and Mike~Del Balso,
\newblock ``Horovod: fast and easy distributed deep learning in {TensorFlow},''
\newblock {\em ArXiv}, vol. abs/1802.05799, 2018.

\bibitem{Infra2019}
C.~Kim, S.~Kim, K.~Kim, M.~Kumar, J.~Kim, K.~Lee, C.~Han, A.~Garg, E.~Kim,
  M.~Shin, S.~Singh, L.~Heck, and D.~Gowda,
\newblock ``{End-to-end training of a large vocabulary end-to-end speech
  recognition system},''
\newblock in {\em 2019 IEEE Automatic Speech Recognition and Understanding
  Workshop (ASRU)}, 2019,
\newblock accepted.

\bibitem{Povey_ASRU2011}
Daniel Povey, Arnab Ghoshal, Gilles Boulianne, Lukas Burget, Ondrej Glembek,
  Nagendra Goel, Mirko Hannemann, Petr Motlicek, Yanmin Qian, Petr Schwarz, Jan
  Silovsky, Georg Stemmer, and Karel Vesely,
\newblock ``The kaldi speech recognition toolkit,''
\newblock in {\em IEEE 2011 Workshop on Automatic Speech Recognition and
  Understanding}. Dec. 2011, IEEE Signal Processing Society,
\newblock IEEE Catalog No.: CFP11SRW-USB.

\bibitem{Sak2015FastAA}
Hasim Sak, Andrew~W. Senior, Kanishka Rao, and Françoise Beaufays,
\newblock ``Fast and accurate recurrent neural network acoustic models for
  speech recognition,''
\newblock {\em ArXiv}, vol. abs/1507.06947, 2015.

\bibitem{tensorflow2015-whitepaper}
Mart\'{\i}n Abadi, Ashish Agarwal, Paul Barham, Eugene Brevdo, Zhifeng Chen,
  Craig Citro, Greg~S. Corrado, Andy Davis, Jeffrey Dean, Matthieu Devin,
  Sanjay Ghemawat, Ian Goodfellow, Andrew Harp, Geoffrey Irving, Michael Isard,
  Yangqing Jia, Rafal Jozefowicz, Lukasz Kaiser, Manjunath Kudlur, Josh
  Levenberg, Dandelion Man\'{e}, Rajat Monga, Sherry Moore, Derek Murray, Chris
  Olah, Mike Schuster, Jonathon Shlens, Benoit Steiner, Ilya Sutskever, Kunal
  Talwar, Paul Tucker, Vincent Vanhoucke, Vijay Vasudevan, Fernanda Vi\'{e}gas,
  Oriol Vinyals, Pete Warden, Martin Wattenberg, Martin Wicke, Yuan Yu, and
  Xiaoqiang Zheng,
\newblock ``{TensorFlow}: Large-scale machine learning on heterogeneous
  systems,'' 2015,
\newblock Software available from tensorflow.org.

\bibitem{tflite}
Google Inc.,
\newblock ``{Tensorflow Lite},'' Online documents;
  \url{https://www.tensorflow.org/lite}, 2018,
\newblock [Online; accessed 2018-02-07].

\end{thebibliography}

\end{document}